\documentclass[conference]{IEEEtran}
\IEEEoverridecommandlockouts

\usepackage{cite}
\usepackage{amsmath,amssymb,amsfonts}
\usepackage{algorithmic}
\usepackage{graphicx}
\usepackage{textcomp}
\usepackage{xcolor}
\usepackage{multirow}
\usepackage{tabularx}
\usepackage{enumitem}
\usepackage{soul}
\usepackage{booktabs}
\usepackage{subcaption}
\usepackage{pgfplots}
\usepackage{filecontents}
\usepackage{algorithm}
\usepackage[none]{hyphenat}
\usepackage{balance}
\usepackage{tikz}
\usepackage{graphicx} 
\usetikzlibrary{shapes.geometric, arrows, fit, positioning}

\tikzstyle{startstop} = [rectangle, rounded corners, minimum width=3cm, minimum height=1cm,text centered, draw=black]
\tikzstyle{process} = [rectangle, minimum width=3cm, minimum height=1cm, text centered, draw=black]
\tikzstyle{arrow} = [thick,->,>=stealth]
\def\BibTeX{{\rm B\kern-.05em{\sc i\kern-.025em b}\kern-.08em
    T\kern-.1667em\lower.7ex\hbox{E}\kern-.125emX}}
\begin{document}

\title{A Robust and Efficient Pipeline for Enterprise-Level Large-Scale Entity Resolution}

\author{
\IEEEauthorblockN{
Sandeepa Kannangara\IEEEauthorrefmark{1},
Arman Abrahamyan\IEEEauthorrefmark{1},
Daniel Elias\IEEEauthorrefmark{1},
Thomas Kilby\IEEEauthorrefmark{2},\\
Nadav Dar\IEEEauthorrefmark{3},
Luiz Pizzato\IEEEauthorrefmark{1},
Anna Leontjeva\IEEEauthorrefmark{1},
Dan Jermyn\IEEEauthorrefmark{1}
}
\IEEEauthorblockA{\IEEEauthorrefmark{1}Commonwealth Bank of Australia, Sydney, Australia \\
\{sandeepa.kannangara, arman.abrahamyan, daniel.elias, luiz.pizzato1, anna.leontjeva, dan.jermyn\}@cba.com.au}
\thanks{\IEEEauthorrefmark{2}Affiliated with Commonwealth Bank of Australia at the time of research. Current contact: \texttt{thomas.kilby1102@gmail.com}}
\thanks{\IEEEauthorrefmark{3}Affiliated with Commonwealth Bank of Australia at the time of research. Current contact: \texttt{nadavdar2@gmail.com}}
}

\maketitle

\begin{abstract}
Entity resolution (ER) remains a significant challenge in data management, especially when dealing with large datasets. This paper introduces MERAI (Massive Entity Resolution using AI), a robust and efficient pipeline designed to address record deduplication and linkage issues in high-volume datasets at an enterprise level. The pipeline's resilience and accuracy have been validated through various large-scale record deduplication and linkage projects. To evaluate MERAI's performance, we compared it with two well-known entity resolution libraries, Dedupe and Splink. While Dedupe failed to scale beyond 2 million records due to memory constraints, MERAI successfully processed datasets of up to 15.7 million records and produced accurate results across all experiments. Experimental data demonstrates that MERAI outperforms both baseline systems in terms of matching accuracy, with consistently higher F1 scores in both deduplication and record linkage tasks. MERAI offers a scalable and reliable solution for enterprise-level large-scale entity resolution, ensuring data integrity and consistency in real-world applications.
\end{abstract}


\section{Introduction}

Entity resolution (ER) represents a fundamental problem in data management, encompassing the identification of records that refer to identical real-world entities across heterogeneous data sources. Also referred to as entity matching, record linkage~\cite{doi:10.1080/01621459.1969.10501049}, reference reconciliation~\cite{10.1145/1066157.1066168}, and merge-purge~\cite{10.1145/223784.223807}, ER addresses the challenge of establishing entity correspondence in the absence of universal identifiers. The significance of this problem has increased substantially with the proliferation of distributed data systems and the exponential growth of enterprise data volumes.

Contemporary enterprise environments present particularly challenging conditions for ER implementation. Organisations typically maintain multiple data systems across business units, often including legacy applications with heterogeneous data schemas and inconsistent entity representations. The Commonwealth Bank of Australia (CBA), managing data for approximately 17 million customers~\cite{cbaweb}, exemplifies these challenges at scale. The institution's data infrastructure encompasses numerous digital platforms, necessitating sophisticated ER capabilities to maintain data consistency and support analytical processes.

The enterprise ER problem is characterised by several critical requirements that distinguish it from traditional academic formulations. Processing datasets containing tens of millions of records requires computational complexity that scales linearly rather than quadratically with dataset size. Enterprise applications demand precision and recall levels suitable for high-stakes environments where entity misidentification carries significant operational and regulatory consequences. These systems must complete large-scale deduplication and linkage operations within operationally acceptable time constraints.

Despite extensive academic research in ER~\cite{10.5555/2344108}, current solutions remain inadequate for practical enterprise deployment. According to Barlaug~\cite{barlaug2020tailoring}, most industrial deduplication and record linkage continues to be performed ad-hoc on a case-by-case basis, failing to leverage published ER research. They argue that this gap exists primarily due to limited production readiness, as openly accessible, well-documented, production-ready libraries developed from high-quality ER research are rare. The authors further observe that organisations exhibit reluctance to invest significantly in developing and integrating ER solutions due to increased failure risk.

Additional challenges arise from scalability deficiencies, where most available solutions are not designed to perform ER on large datasets, exhibiting poor performance when scaled to millions or billions of records. This limitation becomes more pronounced as datasets grow increasingly complex and heterogeneous. Current ER systems also suffer from incomplete pipeline coverage, typically supporting only specific pipeline components such as indexing and matching~\cite{10.14778/2994509.2994535}, while neglecting critical stages such as clustering.

These empirical findings demonstrate a substantial gap between academic ER research and practical enterprise requirements, motivating the development of MERAI (Massive Entity Resolution using AI). MERAI addresses these limitations through novel algorithmic optimisations within an architecture designed to satisfy enterprise-scale demands for accuracy, scalability, and reliability. Following its implementation, MERAI was widely adopted across the bank for executing ER projects.

This research contributes to the field through systematic characterisation of enterprise-scale ER requirements and empirical analysis of existing solution limitations. The work presents improved algorithms through the development of optimised blocking and clustering techniques that achieve linear computational complexity while maintaining high accuracy. The paper describes the design and implementation of a complete ER pipeline encompassing the full workflow from data preprocessing through indexing, matching, and clustering stages, capable of processing datasets at enterprise scale. Additionally, detailed empirical evaluation demonstrates strong performance compared to state-of-the-art frameworks including Dedupe~\cite{dedupe2022} and Splink~\cite{Linacre_Lindsay_Manassis_Slade_Hepworth_2022} across multiple evaluation metrics.

The paper is structured as follows: Section~\ref{sec:er} provides technical background on ER methodology and reviews existing solutions. Section~\ref{sec:approach} presents the MERAI system architecture and detailed algorithmic specifications. Section~\ref{sec:evaluation} reports experimental results from comparative evaluations with established ER frameworks across datasets of varying scale and complexity.

\section{Entity Resolution}
\label{sec:er}
ER encompasses the systematic identification of records that refer to identical real-world entities within or across data sources when unique identifiers are unavailable. The ER process operates on structured data records, addressing both deduplication (within single datasets) and record linkage (across multiple datasets)~\cite{10.1145/3418896}.

The standard ER workflow comprises three sequential stages, as illustrated in Figure~\ref{fig:entity-resolution}~\cite{dong2013big,10.5555/2344108}:

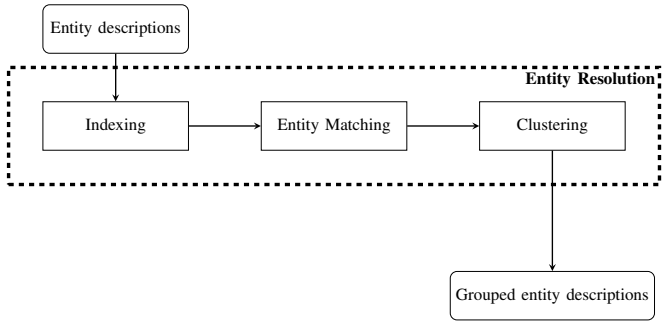
\begin{figure}[h!]
\centering
\resizebox{\columnwidth}{!}{ 
	\begin{tikzpicture}[node distance=2cm]

    \node (start) [startstop] {Entity descriptions};
    \node (indexing) [process, below of=start] {Indexing};
    \node (matching) [process, right of=indexing,xshift=2.5cm] {Entity Matching};
    \node (clustering) [process, right of=matching,xshift=2.5cm] {Clustering};
    \node (end) [startstop, below of=clustering,yshift=-1.5cm] {Grouped entity descriptions};
    
    \draw[arrow] (start) -- (indexing);
    \draw[arrow] (indexing) -- (matching);
    \draw[arrow] (matching) -- (clustering);
    \draw[arrow] (clustering) -- ++(0,-1.5cm)-|  node[pos=.25,left]{ }(end);
    
    \node[draw, dashed, line width=0.7mm, fit=(indexing)(matching)(clustering), inner sep=0.7cm, name=groupbox] {};
    
    \node[anchor=north east] at (groupbox.north east) {\textbf{Entity Resolution}};

  \end{tikzpicture}
}
\caption{Entity Resolution Workflow}
\label{fig:entity-resolution}
\end{figure}

\subsection{ER Workflow Components}
\subsubsection{Indexing}
The technique of indexing, also known as blocking, tackles the computational burden created by exhaustive pairwise comparisons. For datasets of size $n$, naive comparison requires $(n^2-n)/2$ operations, resulting in quadratic complexity~\cite{10.5555/2344108}. Indexing techniques group potentially matching records into blocks based on shared characteristics, ensuring comparisons occur only between candidates within common blocks~\cite{10.1145/3377455}. This approach considerably reduces computational requirements while maintaining recall through careful selection of blocking criteria.

\subsubsection{Entity Matching}
The matching stage determines whether candidate pairs identified during indexing refer to identical entities. Contemporary approaches employ various methodologies including string similarity measures, probabilistic models, supervised learning, unsupervised learning, and rule-based systems~\cite{papadakis2021four}. Machine learning-based techniques demonstrate superior accuracy by learning matching probabilities from training examples~\cite{10.1145/3418896}. Recent research indicates that learning-based approaches consistently outperform rule-based methods in practical applications~\cite{10.14778/2994509.2994535,mudgal2018deep}.

\subsubsection{Entity Clustering}
Clustering transforms pairwise matching decisions into entity groups representing distinct real-world objects~\cite{hassanzadeh2009framework,nentwig2016holistic}. This stage operates on a similarity graph where nodes represent records and weighted edges indicate matching likelihood. Clustering algorithms can identify transitive relationships, grouping entities that lack direct pairwise connections but share common intermediary matches. The output consists of disjoint clusters, each containing all records corresponding to a single real-world entity~\cite{10.1145/3418896}.

\subsection{Existing ER Solutions}

Several open-source and commercial ER tools are available for structured data processing~\cite{10.14778/2994509.2994535,10.1145/3418896,binette2022almost}. This section reviews representative solutions that informed our system design decisions:

\subsubsection{FEBRL}
The Freely Extensible Biomedical Record Linkage (FEBRL) library~\cite{10.1145/1401890.1402020} provides a graphical user interface incorporating both the Fellegi-Sunter probabilistic framework and various supervised classification algorithms. FEBRL includes basic clustering functionality but demonstrates several limitations for enterprise deployment: its primary focus on healthcare domain applications, limited scalability validation on large datasets, and lack of comprehensive clustering capabilities for general-purpose ER tasks.

\subsubsection{Magellan}
Magellan~\cite{Govind2019} offers a comprehensive development framework for ER model construction, supporting both rule-based systems and supervised machine learning classifiers. A distinguishing feature is its Deep Learning module, which represents a unique capability among available ER tools. However, Magellan's architecture lacks integrated clustering functionality for processing matching pairs, requiring external solutions to complete the full ER pipeline.

\subsubsection{Dedupe}
Dedupe~\cite{dedupe2022} combines the Fellegi-Sunter~\cite{fellegi1969theory} probabilistic record linkage method with active learning to continually train itself using examples from the data. The approach allows Dedupe to choose optimal threshold weights to differentiate between matches and non-matches. Dedupe also has the ability to cluster similar records together. The clustering is done using hierarchical agglomerative clustering with a centroid linkage. Generated clusters of linked records are provided with confidence scores, which help estimate the measure of certainty for each linkage.

Dedupe is an excellent tool that can be easily used to identify and eliminate duplicate records within datasets, enhancing data quality and ensuring more accurate analysis. However, we observed that it does not scale well with large datasets, preventing us from using it with millions of records. The Dedupe library's documentation\footnote{https://docs.dedupe.io/en/latest/Troubleshooting.html} mentions memory bottlenecks when working with large datasets (larger than 10K records). We encountered these memory limitations when testing Dedupe on datasets containing over 3 million records.

\subsubsection{Splink}
Splink~\cite{Linacre_Lindsay_Manassis_Slade_Hepworth_2022} represents a more recent addition to open-source ER tools, designed specifically for scalable probabilistic record linkage. The framework implements the Fellegi-Sunter model~\cite{fellegi1969theory} with optimisations for large dataset processing. However, our empirical evaluation revealed significant accuracy deficiencies that render Splink unsuitable for high-precision enterprise applications, particularly in domains where entity misidentification carries substantial operational consequences.

\subsection{Gap Analysis}
The reviewed solutions demonstrate consistent limitations that prevent their adoption in enterprise-scale applications. Most frameworks exhibit poor performance characteristics when processing datasets containing millions of records, with memory allocation patterns and algorithmic complexity preventing linear scaling. Existing solutions typically optimise for either computational efficiency or matching accuracy, failing to achieve both objectives simultaneously. Additionally, many tools provide only partial ER pipeline coverage, requiring integration of multiple systems to achieve end-to-end functionality, which increases system complexity and maintenance overhead.

These identified limitations collectively demonstrate the need for an integrated approach that addresses scalability, accuracy, completeness, and operational readiness simultaneously, providing the foundation for MERAI's design requirements and architectural decisions detailed in the following section.

\section{MERAI Pipeline}
\label{sec:approach}
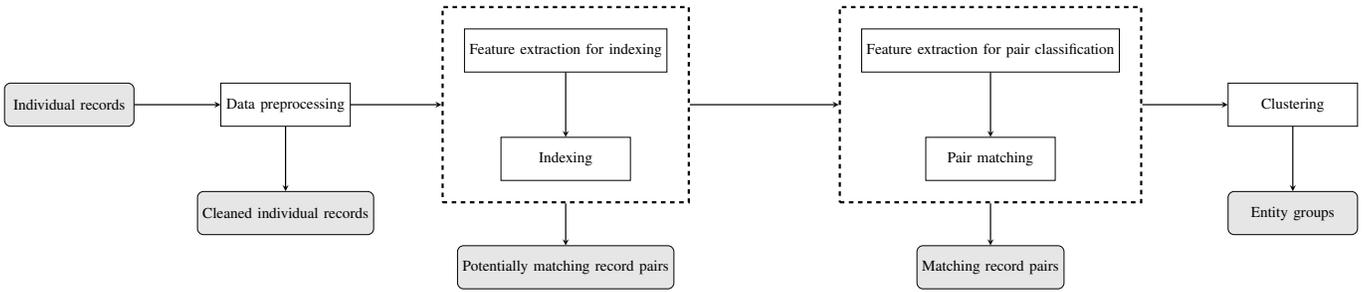
\begin{figure*}[h!]
\centering
\resizebox{\textwidth}{!}{ 
\begin{tikzpicture}[node distance=1.5cm and 2cm]

\node (feature1) [process] {Feature extraction for indexing};
\node (indexing) [process, below=of feature1] {Indexing};
\node[draw, dashed, line width=0.5mm, fit=(feature1)(indexing), inner sep=0.5cm, name=box1] {};
\node[anchor=north west] at (box1.north west) {};

\node (start) [startstop, left=of box1.center, xshift=-8cm, fill=gray!20] {Individual records};
\node (preprocessing) [process, right=of start] {Data preprocessing};

\node (feature2) [process, right=of feature1, xshift=2.5cm] {Feature extraction for pair classification};
\node (pair_matching) [process, below=of feature2] {Pair matching};
\node[draw, dashed, line width=0.5mm, fit=(feature2)(pair_matching), inner sep=0.5cm, name=box2] {};
\node[anchor=north west] at (box2.north west) {};

\node (clustering) [process, right=of box2.center, xshift=3.5cm] {Clustering};

\node (cleaned) [startstop, below=of preprocessing, fill=gray!20] {Cleaned individual records};
\node (potentially_matching) [startstop, below=of indexing, fill=gray!20] {Potentially matching record pairs};
\node (matching_pairs) [startstop, below=of pair_matching, fill=gray!20] {Matching record pairs};
\node (entity_groups) [startstop, below=of clustering, fill=gray!20] {Entity groups};

\draw[arrow] (start) -- (preprocessing);
\draw[arrow] (preprocessing) -- (cleaned);
\draw[arrow] (preprocessing.east) -- (box1.west);
\draw[arrow] (feature1) -- (indexing);
\draw[arrow] (box1.east) -- (box2.west);
\draw[arrow] (feature2) -- (pair_matching);
\draw[arrow] (box2.east) -- (clustering.west);
\draw[arrow] (box1) -- (potentially_matching);
\draw[arrow] (box2) -- (matching_pairs);
\draw[arrow] (clustering) -- (entity_groups);

\end{tikzpicture}
}
\caption{MERAI pipeline for record deduplication and linkage}
\label{fig:pipeline}
\end{figure*}

MERAI is a data deduplication and record linkage solution designed to scale for large, enterprise-level datasets. The MERAI pipeline detailed in Figure~\ref{fig:pipeline} is implemented in Python. It encompasses all steps illustrated in Figure~\ref{fig:entity-resolution} for an ER pipeline. 

MERAI begins with a set of individual data records as input. To enable efficient parallel processing due to the large scale of the data, the input data is first partitioned into smaller chunks at the column level. For deduplication, a single table is used, while record linkage uses two tables. The following subsections describe each step of the pipeline in more detail.

\subsection{Data preprocessing}

MERAI begins by generating a data profiling report that summarises key statistics for each column, including the distribution of the counts of unique values, the number of null values, the number of unique values, and a sample of the most frequently occurring values. This profiling report serves as a crucial resource for identifying data quality issues within the original dataset. The information from the report then guides the subsequent data cleaning and normalisation step, which employs regular expressions to rectify the detected anomalies and inconsistencies.

Once data quality issues are identified through the data profiling report and visual inspection, a set of regular expressions is formulated to clean and normalise the data. These regular expressions are devised to address irregularities and inconsistencies within the data and allow finding, replacing, or removing the detected problematic patterns. For example, a phone number field with dummy values such as `0000000`, `999999999`, `-`, etc. is removed using regular expression matching. The data profiling report can be recomputed to assist in finding new anomalies and iterating to refine regular expressions to enhance the comprehensiveness of the data cleaning process. Finally, columns containing constant values were removed because they offer no variability to assist in the matching process. 

\subsection{Indexing}

The indexing step in MERAI is designed to identify potential matching pairs for pairwise matching. The indexing algorithm developed for MERAI is a variation of the standard blocking algorithm~\cite{fellegi1969theory,10.5555/2344108}. Standard blocking is a foundational technique in record linkage and deduplication that efficiently reduces the number of candidate pairs for comparison. It works by grouping records into blocks based on shared attributes, allowing comparisons only within the same block. 

To address the scalability and accuracy requirements of enterprise-level ER, we enhanced the standard blocking algorithm with three key modifications:
\begin{enumerate}
  \item Use all combinations of selected attributes when creating groups to explore all possible groupings
  \item Save attributes as integers to reduce memory usage
  \item Apply maximum row limitation to ignore large blocks that hinder performance
\end{enumerate}

These modifications work together to create an effective indexing strategy that balances recall, computational efficiency, and memory optimisation. The following subsections detail the implementation and rationale for each enhancement.

\subsubsection{Feature extraction for indexing}\label{index_features}

The foundation of our enhanced indexing approach lies in effective feature extraction that accommodates the complexity of enterprise data structures. To begin the indexing process, we must first identify a set of attributes that can serve as indexing features for each dataset. These attributes are key characteristics that can help distinguish and match records, such as a person's name, address, and date of birth. 

Our approach recognises that enterprise datasets often contain heterogeneous attribute representations, leading us to categorise attributes into two fundamental types that enable flexible matching strategies:

\textbf{Scalar attributes} contain only a single column for the attribute in the dataset. For instance, if a dataset has only one column for date of birth, it would be considered a scalar attribute. Scalar attributes are straightforward to compare, as there is only one value per record, enabling direct one-to-one comparisons.

\textbf{List attributes} have multiple columns for the same attribute type. For example, if a dataset contains two columns for addresses (e.g., ``Registered Address'' and ``Residential Address''), the address attribute would be classified as a list attribute. By treating such attributes as lists, we can compare a value from one record to any value within the list of another record, rather than strictly comparing within a single column. This flexibility enhances the chances of finding matches, even if the values are in different columns across records.

This categorisation enables comprehensive attribute combination generation by recognising semantic relationships between columns, allowing effective blocking strategies that capture entity variations across different data representations. The practical implementation constructs an expanded table where each row represents a unique combination of values from list columns. Table~\ref{index_feature} illustrates an example with two addresses and three phone numbers, resulting in six expanded rows representing all address-phone combinations for extensive comparison coverage.

\begin{table}  
	\caption{Example representation of one single record with one scalar value (date of birth - DOB), and two list types (Phone No and Address) represented as an expanded six rows in the database}
	\begin{tabularx}{\linewidth}{p{.15\linewidth}|>{\raggedleft\arraybackslash}p{.2\linewidth}|>{\raggedleft\arraybackslash}p{.2\linewidth}|>{\raggedleft\arraybackslash}p{.25\linewidth}}
	 \toprule
	 Row ID & DOB & Phone No &  Address \\  
	 \midrule    
	 10001 & 1978-03-19 & 0511111111 & 2 Acadaca St Sydney 2000 \\
        10001 & 1978-03-19 & 0533333333 & 2 Acadaca St Sydney 2000 \\
        10001 & 1978-03-19 & 0599999999 & 2 Acadaca St Sydney 2000 \\
        10001 & 1978-03-19 & 0511111111 & 4 Down Under Rd Perth 6000 \\
        10001 & 1978-03-19 & 0533333333 & 4 Down Under Rd Perth 6000 \\
        10001 & 1978-03-19 & 0599999999 & 4 Down Under Rd Perth 6000 \\
        \bottomrule    
    \end{tabularx}
    \label{index_feature}
\end{table}

To implement our second modification for memory optimisation, the actual attribute values in the expanded table are replaced with integer mappings. Each unique value within a column is assigned a corresponding integer, and these mappings are stored separately. This technique reduces the memory footprint while still allowing for efficient value-based matches, directly addressing the scalability constraints identified in existing solutions.

\subsubsection{Indexing process}

The indexing process generates all possible attribute combinations to ensure extensive coverage of potential matching scenarios. Taking the attribute set defined in Section~\ref{index_features} as input, the main objective is to divide records into groups based on attribute subsets and generate all possible pairs within each group. Our enhanced approach generates multiple attribute set combinations systematically, ensuring that records with partial attribute matches are not overlooked. For instance, using the example in Table~\ref{index_feature}, records can be grouped by phone number, address, or combinations thereof, with all possible pairs generated within each group.

Our third modification addresses computational challenges from highly common attribute values by employing configurable group size limits. Attribute sets such as country, last names, or date of birth may create large groups due to value commonality, undermining computational efficiency. The optimal maximum group size varies based on the domain and nature of the dataset—for customer profile datasets, a threshold of 1,000 records has proven effective in our experiments. This mechanism balances computational feasibility with matching completeness, maintaining linear scaling properties for multi-million record datasets. The complete indexing approach is formalised in Algorithm~\ref{algo:indexing}. 

Algorithm~\ref{algo:indexing} has time complexity $O(2^F + 2^F \cdot n + 2^F \cdot G \cdot \text{maxrow}^2)$. Here, $n$ is the number of records, $F$ is the number of features, $G$ is the average number of groups per feature combination, and $\text{maxrow}$ is the maximum allowed group size. The exponential dependence on feature count necessitates keeping $F$ small in practice, while the $\text{maxrow}$ parameter prevents quadratic blow-up in pair generation for common attribute values. For typical entity resolution datasets with skewed value distributions, the algorithm achieves efficient performance by focusing pair generation only on moderately-sized groups while maintaining recall through feature combination coverage.

\begin{algorithm}
	\caption{Indexing process}
	\label{algo:indexing}
	\begin{algorithmic}
		\STATE \textbf{Input}: DataFrame $T$, List of features $F$, integer $maxrow$
		\STATE \textbf{Output}: List of pairs of records that meet the criteria
		\STATE \textbf{Steps}:
		\STATE Initialise an empty list $combinations\_list$
		\FOR{each integer $i$ from 1 to the length of features:}
		\STATE Generate all combinations of $features$ of length $i$
		\STATE Append each combination to $combination\_list$
		\ENDFOR
		\STATE Initialise an empty list $included\_pairs$
		\FOR{each $feature\_subset$ in $combination\_list$:}
		\STATE Group $T$ by $feature\_subset$.
		\FOR{each group:}
		\IF{the group has $maxrow$ or fewer rows:}
		\STATE Get all the record pair combinations within the group
		\STATE Add the record pairs to $included\_pairs$
		\ENDIF
		\ENDFOR
		\ENDFOR
		\STATE \textbf{return} $included\_pairs$
	
	\end{algorithmic}
\end{algorithm}

\subsection{Matching}
The indexing step generates an extensive list of candidate record pairs requiring further evaluation to determine actual matches. To achieve this refinement, we employ supervised machine learning classification techniques that learn to identify matching patterns from labelled training data. This section details our machine learning-based approach for accurate pair classification and match probability estimation.

\subsubsection{Feature extraction for pair classification}

To train a machine learning model, we generate features from all record pairs in the indexing output using string similarity-based numeric features for each attribute. For every possible record match $\langle a,b \rangle$, we compare corresponding attributes, such as $(address\_a, address\_b)$, and create features like $\delta\_address(a, b)$ where $\delta$ represents a similarity function.

For feature generation, MERAI utilises the Magellan package~\cite{10.14778/2994509.2994535}, which provides two categories of similarity functions:

\begin{itemize}

      \item Token-based similarity functions: These functions apply a tokenizer to split each string into tokens and compute the similarity between the two token sets. The token-based similarity fuctions in Magellan package are Overlap coefficient, Dice, Cosine and Jaccard similarity.

      \item Non-token-based similarity functions: These functions directly compute the similarity between two strings without the need for a tokenizer. They calculate the number of operations required to transform one string into another, with a higher number of operations indicating less similarity. The non-token-based similarity functions included in Magellan are Levenshtein similarity/distance, Jaro, Exact match, Jaro-Winkler, Needleman-Wunsch, Smith-Waterman, Monge-Elkan and Absolute Norm.      

\end{itemize}

Magellan applies predefined heuristic rules for feature generation, selecting appropriate similarity functions based on attribute data types determined by average word tokens or numeric/boolean classification. However, this rule-based approach has limitations~\cite{wang2021automating}, particularly in similarity function selection based on average word count, which ignores data quality issues. For instance, attributes with average word counts exceeding 10 are limited to Cosine and Jaccard distances, even though shorter strings within those attributes may benefit from other functions.

To address this limitation, we adopted the solution proposed by~\cite{wang2021automating} of applying all similarity functions regardless of string length, while maintaining Magellan's feature set for numeric and boolean data types.

\subsubsection{Pair matching}

Using potential matching pairs identified in the indexing step, this step aims to probabilistically match those pairs. MERAI uses supervised machine learning to classify record pairs as matches or non-matches. The model is trained on the text similarity-based feature vector discussed in the previous section.  

To train a supervised machine learning model, it is essential to have labelled data. The process involves randomly selecting some records, followed by human annotators labelling each record pair as a match or not. Then, we tested a number of machine learning algorithms to understand which would perform the best for this problem. We tested decision trees, Naive Bayes, SVM, XGBoost and random forests.  

The trained model is then applied to the remaining unlabelled data to compute the probability of a pair of records being a match. We used methods that output match probabilities, allowing us to adjust the level of certainty required when matching records. For tasks where false positives are unacceptable, we can ensure only matches with high probabilities are accepted, thereby maintaining the accuracy and reliability of the results.

\subsection{Clustering}

Probabilistic matching identifies record pairs referring to the same entity, but entities may appear across multiple pairs. This step identifies all connected records belonging to the same entity.

Matched pairs create a graph where nodes represent entity records and edges indicate matching pairs. Graph theory enables connecting related nodes through \textit{connected components}, forming complete subgraphs called \textit{cliques}. A clique contains nodes where each connects to all others within the subset. For example, if records A-B and A-C are matched, graph theory establishes the A-B-C connection to label them as a single entity.

\begin{figure}
	\includegraphics[width=\linewidth]{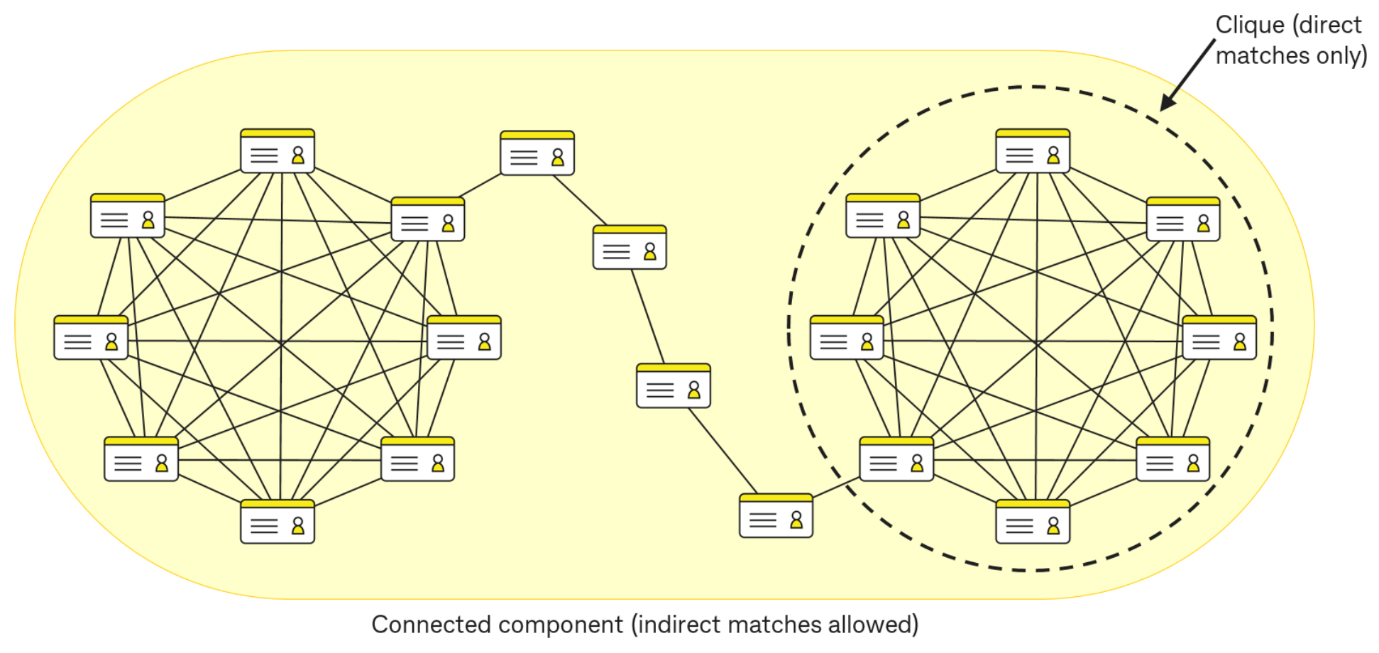}
	\caption{Cliques within a connected component}
	\label{fig:connected_components}
\end{figure}

Complex scenarios involve many connected records with overlapping cliques, as shown in Figure~\ref{fig:connected_components}. The algorithm identifies both connected components and constituent cliques within them.

The entity identification process follows these steps:

\begin{enumerate} 

    \item Graph construction 

    \begin{enumerate} 

        \item Edges: Pairs of entities estimated to be matches.

        \item Edge weights: Match probabilities obtained from the classification model.

        \item Nodes: Entities involved in edges.

    \end{enumerate} 

    \item Identification of connected components within the constructed graph.

    \item Identification of disjoint cliques within the connected components. 

    \begin{enumerate} 

        \item Identification of maximally-sized cliques (these cliques may overlap in general).

        \item Adjustment of the identified cliques to make them non-overlapping. Hence a given node is only belongs to one clique.

    \end{enumerate} 

\end{enumerate} 

To ensure the cliques are disjoint (as described in step 3.b), nodes that belonged to multiple cliques identified in step 3.a were assigned to only one clique and removed from the others. Various algorithms have been proposed in the literature for generating disjoint cliques~\cite{gyori1992edge,jansen1997disjoint,yuster2014edge}. We developed a greedy algorithm (Algorithm~\ref{algo:disjointalgorithm}) aimed at maximising the edge weights within the cliques, thereby eliminating low-weight edges.

\begin{algorithm}
    \caption{Disjoint Cliques Generation}
    \label{algo:disjointalgorithm}
    \begin{algorithmic}[1]
        \STATE \textbf{Input:} Graph $G$ (with edge weights $w_e$; if no weights, set all $w_e=1$)
        \STATE \textbf{Output:} List $L$ of disjoint cliques in $G$
        \STATE Initialise an empty list $L$
        \STATE Define helper function $\text{edge\_weight\_loss}(a, b, G)$:
        \begin{ALC@g}
            \STATE \textbf{Parameters:} Sets of vertices $a$ and $b$ in $G$
            \STATE \textbf{Return:} Sum of weights of edges in $G$ joining a vertex in $(a \cap b)$ with a vertex in $(b - a)$
            \STATE $\text{intersection} \gets a \cap b$
            \STATE $\text{difference} \gets b - a$
            \STATE $\text{loss} \gets 0$
            \FOR{each vertex $u$ in $\text{intersection}$}
                \FOR{each vertex $v$ in $\text{difference}$}
                    \IF{$G$ has edge $(u, v)$}
                        \STATE $\text{loss} \gets \text{loss} + \text{weight of edge } (u, v)$
                    \ENDIF
                \ENDFOR
            \ENDFOR
            \STATE \textbf{return} $\text{loss}$
        \end{ALC@g}
        \WHILE{$G$ is not empty}
            \STATE $C \gets \text{list of all sets of vertices in $G$ which represent cliques}$
            \STATE $n \gets \max(\text{length}(s) \text{ for } s \text{ in } C)$
            \STATE $\text{candidates} \gets \text{list of sets in } C \text{ that have length } n$
            \STATE Sort $\text{candidates}$ such that: in each adjacent pair $(a, b)$, $\text{edge\_weight\_loss}(a, b, G) \geq \text{edge\_weight\_loss}(b, a, G)$
            \STATE $\text{candidate}_0 \gets \text{first set in candidates}$
            \STATE Remove all vertices in $\text{candidate}_0$ from $G$
            \STATE Append $\text{candidate}_0$ to $L$
        \ENDWHILE
        \STATE \textbf{Return} $L$
    \end{algorithmic}
\end{algorithm}

\begin{figure}
	\centering
	\begin{subfigure}{0.7\linewidth}
		\includegraphics[width=\linewidth]{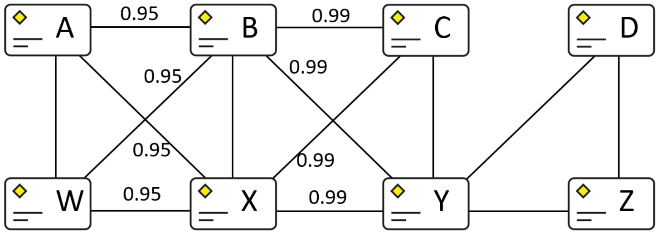}
		\caption{Initial graph with selected edge weights}
		\label{fig:first}
	\end{subfigure}
	\hfill
	\begin{subfigure}{0.7\linewidth}
		\includegraphics[width=\linewidth]{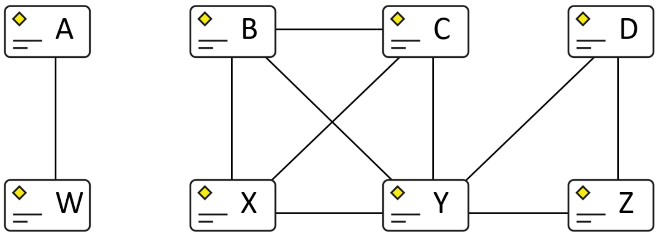}
		\caption{After first iteration}
		\label{fig:second}
	\end{subfigure}
	\hfill
	\begin{subfigure}{0.7\linewidth}
		\includegraphics[width=\linewidth]{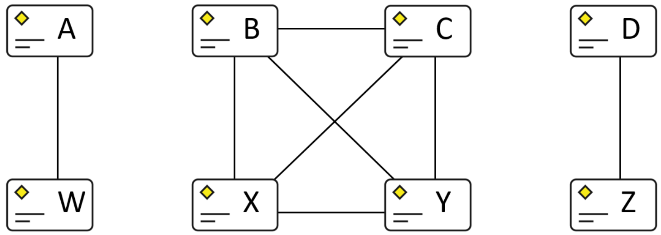}
		\caption{After second iteration (disjoint cliques)}
		\label{fig:third}
	\end{subfigure}
			
	\caption{The process of generating disjoint cliques }
	\label{fig:disjointcliques}
\end{figure}

To illustrate the use of this algorithm, consider the connected component depicted in Figure\ref{fig:first}. In this case, the candidates identified in line 21 of Algorithm\ref{algo:disjointalgorithm} are cliques of the maximum size: ABWX and BCXY. To separate these cliques, the edges AB, WB, AX, and WX are removed, resulting in a total edge weight loss of 3.80 (four edges each with weight 0.95). After removing these edges, the adjusted connected component is shown in Figure\ref{fig:second}. Applying the same procedure as in the first iteration, the candidates are now: {BCXY, DYZ}. The maximum clique size remains four, but there is only one clique of this size (BCXY). Therefore, edges YD and YZ are removed from the smaller clique (DYZ) to produce the final disjoint cliques, as shown in Figure\ref{fig:third}.

\section{Evaluate performance of MERAI}
\label{sec:evaluation}

\subsection{Experimental Setup}

To evaluate MERAI's performance, several experiments were conducted using a large public dataset.
Dedupe~\cite{dedupe2022} and Splink~\cite{Linacre_Lindsay_Manassis_Slade_Hepworth_2022} were used as baselines for comparison. This section discusses the statistics and features of the datasets, the preparation of training and testing data, and provides an introduction to the baselines.

\subsubsection{Evaluation dataset}
\label{selecteddataset}
The MERAI pipeline was initially applied to exceptionally large datasets within the bank, consisting of 33 million records. However, for the purposes of the paper, we used a publicly available large-scale dataset.

One such database is the voter registration database, which is publicly available in the United States. The North Carolina State Board of Elections (NCSBE)\footnote{http://www.ncsbe.gov/} provides regular updates to its voter registration database, making it freely and publicly accessible. This dataset has led to developments in entity resolution as well as open-source software and data sets~\cite{christen2014preparation,wortman2019record}. This dataset contains rich information, such as first and last names, years of birth, phone numbers, and addresses, which is similar to the datasets used in enterprises' entity matching projects.

Three data snapshots (02/11/2010, 03/11/2015, and 03/11/2020)\footnote{https://dl.ncsbe.gov/index.html?prefix=data/Snapshots/} were selected to conduct three deduplication experiments and three record linkage experiments. The statistics of the selected datasets are presented in Table~\ref{dataset}. The nature of voter registration datasets facilitates the performance of both deduplication and record linkage tasks. Within a single data snapshot, duplicate entities for the same person may exist due to various reasons, such as people moving or getting married. Similarly, since the same person can appear in two different snapshots with either the same or different personal information, it is possible to link records across two or more snapshot datasets.

\begin{table}   
   \caption{Statistics of the datasets used for experiments}
   \begin{tabularx}{\linewidth}{p{0.33\linewidth}|r|r|r}
   \toprule
      & 2010 & 2015   & 2020\\
   \midrule
   No. of records & 9,697,808 & 6,676,566 & 15,770,705 \\
   Unique NCIDs & 9,300,301 & 6,310,162 & 13,334,038 \\
   NCIDs with duplicates & 365,807 & 300,473 & 1,789,149\\   
   \end{tabularx}
   \label{dataset}
\end{table}

To evaluate performance, it is essential to have labelled data. The North Carolina identification number (NCID) is a unique identifier for everyone in the state that can be used as ground truth to generate labelled data for model training. However, several studies~\cite{christen2014preparation,wortman2019record} have indicated that the NCID is not a reliable unique identifier. To obtain high-quality data with the largest number of correct entity matches, we employed two methods: (1) the attribute combination similarity-based method proposed by~\cite{christen2014preparation}, and (2) selecting a sample of approximately 10K records where we manually verified every label. These approaches improve the overall quality of the dataset for model training and evaluation.

\begin{table}   
   \caption{Statistics of the generated groundtruth based on NCID and attribute combination similarity}
   \begin{tabularx}{\linewidth}{l|>{\raggedleft\arraybackslash}p{0.18\linewidth}|>{\raggedleft\arraybackslash}p{0.18\linewidth}|>{\raggedleft\arraybackslash}p{0.18\linewidth}}
   \toprule
      Dataset & NCID based pairs & Attribute comb. based pairs   & Total no. of pairs\\
   \midrule
   2010 dedup. & 432,042 & 626,031 & 1,246,359 \\
   2015 dedup. & 440,577 & 478,628 & 1,032,881 \\
   2020 dedup. & 3,279,841 & 1,289,360 & 6,417,883\\
   2010-2015 link. & 6,224,570 & 6,108,861 & 11,540,019\\
   2010-2020 link. & 11,930,590 & 10,844,473 & 29,953,436\\
   2015-2020 link. & 8,897,206 & 7,632,214 & 19,972,940 \\
   \end{tabularx}
   \label{groundtruth}
\end{table}

We selected a sample of approximately 10K records from each dataset and manually verified them to build the ground truth data. To obtain the highest quality dataset possible, we first ran MERAI on the full dataset. The indexing output of MERAI was then used to randomly select 10K pairs. This approach increases the likelihood of obtaining matching pairs while keeping the ratio relatively low (see Table~\ref{sampledata}). Each dataset was subsequently split into train test as 70/30. Finally, the selected training and test sets were used to evaluate the performance of both MERAI and and selected baselines.

\begin{table}   
   \caption{Statistics of the sample of data used for performance evaluation}
   \begin{tabularx}{\linewidth}{p{0.25\linewidth}|>{\raggedleft\arraybackslash}p{0.2\linewidth}|>{\raggedleft\arraybackslash}p{0.2\linewidth} | >{\raggedleft\arraybackslash}p{0.1\linewidth}}
   \toprule
      Dataset & Matching pairs \% in training (7k) & Matching pairs \% in test (3k) & Unique records\\
   \midrule
   2010 dedup.   & 1.3\% & 1.4\% & 13,982 \\
   2015 dedup. & 1.3\% & 0.9\% & 13,988 \\
   2020 dedup. & 2.0\% & 2.1\% & 13,994\\
   2010-2015 link. & 5.4\% & 5.3\% & 13,992\\
   2010-2020 link. & 6.3\% & 6.0\% & 13,995\\
   2015-2020 link. & 6.8\% & 6.7\% & 13,993 \\
   \end{tabularx}
   \label{sampledata}
\end{table}

\subsubsection{Dedupe as a baseline}
\label{sec:dedupe}

Dedupe~\cite{dedupe2022} performs pairwise matching through an active learning process, where the user is given pairs of rows and instructed to label whether they are a match or not. The package requires at least 10 positive matches and 10 negative matches to be supplied to train a successful model. Once the active learning is completed, the matching and clustering processes are performed without the need for user input. 

Even though the typical Dedupe pipeline uses active learning, it is feasible to simulate human input by feeding Dedupe the labels it requests from ground truth data to ensure consistency~\cite{10.1371/journal.pone.0283811}. Hence, following this approach, we automated the active learning process with our training data.

To correctly use training and test data with Dedupe, we followed a two-step process for each dataset. First, we ran Dedupe with the training dataset, providing training labels for active learning. Then, we ran Dedupe again using the test dataset, along with the model settings file from the previous iteration as input.

\subsubsection{Splink as a baseline}
\label{sec:splink}

Splink~\cite{Linacre_Lindsay_Manassis_Slade_Hepworth_2022} offers a flexible framework for defining comparison rules between records and calculating match probabilities based on those comparisons. It also allows defining blocking rules to efficiently reduce the comparison space. For our implementation, we configured specialised name comparisons for personal identifiers. Categorical fields including sex code and birth year were set with exact matching and appropriate term frequency adjustments. Our blocking strategy combined last name, first name, and birth year matches.

We trained the model using the $estimate\_m\_\allowbreak from\_pairwise\_\allowbreak labels()$ method with positive examples from our labelled training data, enabling the system to learn match probabilities directly from known matches. After training with our training dataset, we generated predictions across the entire dataset and extracted the pairs corresponding to our test dataset for performance assessment.

\subsection{Results}

\subsubsection{Performance of pairwise match} 
\label{sec:performance}

\begin{table*}[t]	
   \centering
     \caption{Deduplication and record linkage pairwise matching model performance comparison}
     \begin{tabularx}{0.8\linewidth}{l|r|r|r|r|r|r|r|r|r}
     \toprule
      \multirow{2}{*}{Experiment} & \multicolumn{3}{c|}{Splink} & \multicolumn{3}{c|}{Dedupe} & \multicolumn{3}{c}{MERAI} \\
      & Precision & Recall & F1 & Precision & Recall & F1 & Precision & Recall & F1\\
      \midrule
     2010 deduplication & 70.9\% & 97.5\% & 82.1\% & 100.0\% & 75.0\% & 85.7\% & 100.0\% & 95.0\% & \textbf{97.4\%} \\
     2015 deduplication & 60.0\% & 85.7\% & 70.6\% & 100.0\% & 78.6\% & 88.0\% & 100.0\% & 92.5\% &  \textbf{94.1\%} \\
     2020 deduplication & 76.5\% & 98.4\% & 86.1\% & 92.4\% & 96.8\% & 94.6\% &  100.0\% & 93.7\% & \textbf{96.7\%}\\
     \midrule
     2010-2015 linkage & 89.6\% & 84.9\% & 87.2\% & 73.7\% & 99.4\% & 85.8\% &  96.7\% & 100.0\% & \textbf{98.3\%}\\
     2010-2020 linkage & 94.1\% & 85.2\% & 89.4\% & 73.7\% & 99.4\% & 84.6\% &  96.5\% & 98.2\% & \textbf{97.3\%}\\
     2015-2020 linkage & 88.6\% & 78.3\% & 83.2\% & 71.6\% & 100.0\% & 83.4\% &  92.6\% & 100.0\% & \textbf{96.2\%} \\
     \end{tabularx}
     \label{results}
   \end{table*}

The selected training and test datasets from three snapshots of NCID data were used to evaluate and compare the performance of MERAI and Dedupe. For MERAI, the entire dataset was used for indexing and feature generation steps. Subsequently, model training and testing were conducted using the selected training and test data separately for pairwise classification.

For Dedupe, deduplicated datasets are required as input for record linkage. Therefore, we used the Pandas package's $drop\_duplicates()$ function to remove duplicate records before performing record linkage. In contrast, MERAI and Splink do not require deduplicated datasets as input for record linkage and can handle datasets containing duplicate records.

Table~\ref{results} presents a comparison of pairwise matching performance for Splink, Dedupe, and MERAI across both deduplication and record linkage tasks. For MERAI, the results represent the model's performance in classifying record pairs from indexing output as matches or non-matches. Dedupe's results are derived from its final output file containing records with cluster numbers. We considered all clusters with more than one record to generate pairs, creating all possible combinations within each cluster and comparing them with test data pairs to calculate performance metrics. For Splink, we based the results on final predictions using a match probability threshold of 0.99.

The results demonstrate that MERAI consistently outperforms both Dedupe and Splink in all experiments, achieving the highest F1 scores throughout. In deduplication experiments, MERAI maintains perfect precision (100\%) while achieving high recall (92.5-95\%), identifying matches accurately while capturing most true duplicates. Dedupe similarly achieves perfect or near-perfect precision but shows lower recall (75-96.8\%), missing more true matches. Splink exhibits the lowest precision (60-76.5\%) among the three pipelines while achieving high recall (85.7-98.4\%), suggesting it tends to over-match records.

For record linkage tasks, the performance differences become more pronounced. MERAI demonstrates an excellent balance of precision (92.6-96.7\%) and recall (98.2-100\%), resulting in F1 scores above 96\% across all linkage experiments. Dedupe shows high recall (99.4-100\%) but lower precision (71.6-73.7\%), indicating it captures nearly all true matches but generates many false positives. Splink takes a more balanced approach with moderate to high precision (88.6-94.1\%) and recall (78.3-85.2\%), but falls short of MERAI's overall performance.

These results highlight MERAI's strong ability to maintain high precision while capturing most true matches, making it valuable for applications where both accuracy and completeness are critical.

\subsubsection{MERAI's ability to handle very large datasets}

Next, we evaluated MERAI's runtime performance and scalability. We used six randomly selected subsets ranging from 10K to 2M records from the 2020 NCID dataset to evaluate runtime performance. For these experiments, we compared MERAI specifically with Dedupe. While Splink is designed for large datasets, we excluded it from our runtime and scalability comparisons because its matching accuracy was significantly lower than both MERAI and Dedupe in our earlier experiments (as shown in Table~\ref{results}). Since accuracy is a prerequisite for meaningful deployment in enterprise environments, we focused our computational efficiency comparisons on systems that met minimum accuracy thresholds.

Our runtime experiments revealed that Dedupe cannot efficiently handle datasets above 2M records due to memory constraints during clustering, with processing time increasing dramatically for larger datasets. Based on these findings, we conducted additional scalability experiments using only MERAI on the three complete NCID datasets (9.6M, 6.6M, and 15.7M records) to demonstrate its full capacity to handle large-scale data.

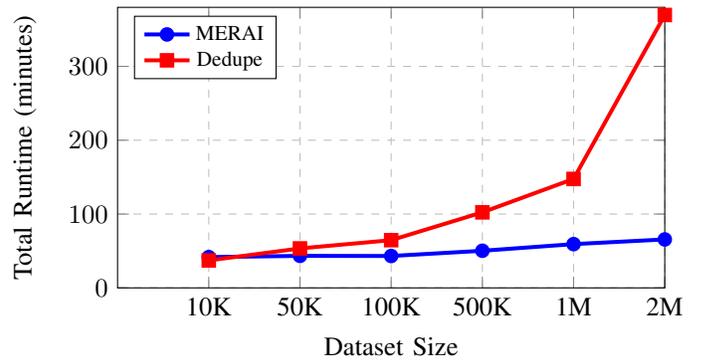
\begin{figure}
    \begin{tikzpicture}
    \begin{axis}[
        width=\linewidth,
        height=0.6\linewidth,
        xlabel={Dataset Size},
        ylabel={Total Runtime (minutes)},
        xmin=0, xmax=6,
        ymin=0, ymax=380,
        xtick={1,2,3,4,5,6},
        xticklabels={10K,50K,100K,500K,1M,2M},
        legend pos=north west,
        grid=both,
        legend style={nodes={scale=0.8, transform shape}},
        title={Comparison of Runtime Performance},
        ymajorgrids=true,
        grid style=dashed,
    ]
    
    \addplot[
        color=blue,
        mark=*,
        line width=1.5pt,
    ]
    coordinates {
        (1,41.72)(2,43.35)(3,43.20)(4,50.27)(5,59.27)(6,65.63)
    };
    \addlegendentry{MERAI}
    
    \addplot[
        color=red,
        mark=square*,
        line width=1.5pt,
    ]
    coordinates {
        (1,37.17)(2,53.45)(3,64.63)(4,102.32)(5,147.55)(6,369.63)
    };
    \addlegendentry{Dedupe}
    
    \end{axis}
    \end{tikzpicture}
    \caption{Comparison of MERAI and Dedupe runtime performance for different dataset sizes}
    \label{fig:runtime}
\end{figure}

\begin{description}[leftmargin=*]
\item[\textbf{Runtime:}]
Table~\ref{time_merai} shows the time taken to complete each step in MERAI and the total time taken. We used 10 CPUs for indexing and 50 for pair classification for all datasets. The results indicate that MERAI's indexing algorithm is highly efficient, with time increasing slowly as dataset size increases. A similar trend can be seen for pair matching and clustering.

We compared MERAI's runtime with Dedupe using the same selected six datasets. We used labelled data files from our performance evaluation experiments in Section\ref{sec:performance} instead of active learning. Table~\ref{time_dedupe} shows the runtime for each step in Dedupe. In the loading and preparation phase, Dedupe reads the data, labels it, and prepares it for training. This preparation runtime increases with dataset size. The model training time is relatively fast for all dataset sizes. However, clustering time dramatically increases above 1M records, likely due to Dedupe's hierarchical clustering algorithm. Traditional hierarchical clustering algorithms typically exhibit a time complexity of $O(n^2)$, making them impractical for large datasets~\cite{10.1109/ieeeconf44664.2019.9048653,10.1142/s0218126619500658}.

Our attempt to run Dedupe on a 3M dataset failed after 24+ hours during clustering due to memory bottlenecks documented in Dedupe's documentation. Hence, we limited the experiments to datasets up to 2M records. As shown in Figure~\ref{fig:runtime}, Dedupe is more efficient for datasets smaller than 10K, while MERAI is more efficient for larger datasets. Dedupe's runtime curve becomes increasingly steep as dataset size grows, while MERAI shows only a minimal increase, making it more suitable for large-scale industrial applications.

\item[\textbf{Scalability:}]
Table~\ref{size} shows MERAI's scalability with the output size it could handle in each step for each experiment with the complete NCID datasets. MERAI successfully handled up to 200 million pairs as indexing output when deduplicating and up to 250 million pairs when record linking. These results demonstrate the effectiveness of MERAI's indexing algorithm. The grouping strategy, which utilises defined features (discussed in Section~\ref{index_features}) and eliminates large groups, improves computational efficiency by replacing total values with unique integer values to make tables more compact.

Furthermore, pairwise feature generation in MERAI effectively processed up to 250 million pairs without failures, providing input for classification. Finally, the scoring output, which includes up to 14 million matching pairs, was successfully used as input to MERAI's clustering algorithm. MERAI uses a graph-based clustering algorithm that can process large datasets without memory bottlenecks. Disjointed cliques are only produced for deduplication to identify unique entities within datasets. For record linkage, clustering was not applied as the objective is to link records between datasets.

By successfully processing datasets of this size, MERAI demonstrates its capability to handle real-world entity resolution tasks involving large-scale data.
\end{description}

\begin{table}  
  \caption{Output of MERAI in each step (values are rounded)}
  \begin{tabularx}{\linewidth}{l|>{\raggedleft\arraybackslash}p{0.15\linewidth}|>{\raggedleft\arraybackslash}p{0.2\linewidth}|>{\raggedleft\arraybackslash}p{0.2\linewidth}}
  \toprule
  & Indexing - Pairs & Classification - Matching pairs & Clustering - Disjoint cliques  \\
  \midrule
  2010 dedup. &118.4M& 1.5M & 1M\\
  2015 dedup. &82.1M& 0.7M & 0.5M\\
  2020 dedup. & 208.7M & 4.4M &  2.3M\\
  2010-2015 link. & 149.3M & 7.6M & N/A\\
  2010-2020 link. & 253.6M &  14.1M & N/A\\
  2015-2020 link. & 177.8M &  10.3M & N/A\\
  \end{tabularx}
  \label{size}
\end{table}

\begin{table}  
   \caption{Time (hh:mm:ss) to complete \textbf{MERAI} pipeline for each step for different dataset sized up to 2M}
   \begin{tabularx}{\linewidth}{p{0.10\linewidth}|>{\raggedleft\arraybackslash}p{0.18\linewidth}|>{\raggedleft\arraybackslash}p{0.18\linewidth}|>{\raggedleft\arraybackslash}p{0.18\linewidth}| >{\raggedleft\arraybackslash}p{0.15\linewidth} }
   \toprule
    Dataset size & Indexing  & Pair classification & Clustering & Total duration\\
   \midrule
   10K & 0:39:26 &   0:02:05  & 0:00:12 &  0:41:43  \\
   50K & 0:38:59 &   0:04:06  & 0:00:16 &  0:43:21  \\
   100K & 0:39:12 & 0:03:45 & 0:00:15  & 0:43:12 \\
   500K & 0:41:27 & 0:08:13 & 0:00:36  & 0:50:16 \\
   1M & 0:43:44 & 0:14:18  & 0:01:44   & 0:59:16   \\
   2M & 0:48:33  & 0:15:46   & 0:01:19   & 1:05:38   \\
   \end{tabularx}
   \label{time_merai}
  \end{table}

  \begin{table}  
   \caption{Time (hh:mm:ss) to complete \textbf{Dedupe} pipeline for each step for different dataset sized up to 2M}
   \begin{tabularx}{\linewidth}{p{0.10\linewidth}|>{\raggedleft\arraybackslash}p{0.18\linewidth}|>{\raggedleft\arraybackslash}p{0.18\linewidth}|>{\raggedleft\arraybackslash}p{0.18\linewidth}| >{\raggedleft\arraybackslash}p{0.15\linewidth} }
   \toprule
    Dataset size & Loading and prep.  & Training & Clustering & Total duration\\
   \midrule
   10K & 0:35:26 &   0:00:23 & 0:01:21 &  0:37:10 \\
   50K & 0:50:26 & 0:00:27 & 0:02:34  & 0:53:27 \\
   100K & 0:53:57 & 0:00:18 & 0:10:23 & 1:04:38 \\
   500K & 0:51:05 & 0:00:30 & 0:50:44 & 1:42:19 \\
   1M & 0:54:07 & 0:00:35 &   1:32:51 &  2:27:33 \\
   2M & 0:57:13 & 0:00:15 & 5:12:10 &  6:09:38  \\
   \end{tabularx}
   \label{time_dedupe}
  \end{table}

\section{Application of MERAI to enterprise level projects}
\label{sec:application}

MERAI was successfully applied to several bank record deduplication and linkage projects, processing up to 33 million records to effectively identify duplicate or matching records across different datasets.

The pipeline consistently handled extensive datasets without any failures in all the projects where it was employed within the bank. MERAI produced final results for each project, demonstrating its robustness and efficiency in processing large-scale data for deduplication and record linkage tasks.

The successful application of MERAI to these real-world projects highlights its reliability and effectiveness in managing and analysing substantial volumes of data. Its ability to handle large datasets and deliver accurate results underscores its suitability for complex entity resolution challenges, not only in banking but also in other industries.

\section{Conclusion and Future Work}
\label{sec:conclusion_future_work}

This paper introduces MERAI, an effective and efficient solution for large-scale enterprise-level entity resolution. The pipeline is designed to handle both record deduplication and record linkage tasks. We validated the robustness and output correctness of the pipeline by applying it to several large-scale record duplication and linkage projects.

We compared MERAI with Dedupe and Splink as baselines to evaluate its performance. The experimental results demonstrate that MERAI outperforms both baselines regarding the accuracy of pairwise classification results. Furthermore, we compared the scalability and runtime of MERAI with Dedupe, showing that MERAI is significantly more robust and scalable. These findings demonstrate MERAI's capability for handling large-scale entity resolution tasks effectively.

It is worth noting that MERAI currently uses features generated by Magellan with slight modifications. An exciting avenue for future work is to design a feature engineering solution tailored explicitly for MERAI, which could potentially further improve its performance.

In conclusion, MERAI presents a robust and efficient approach to enterprise-level entity resolution, capable of handling large-scale datasets and delivering accurate results. Its successful application to real-world projects in the banking industry underscores its potential for addressing complex entity resolution challenges in various domains.

\section*{Acknowledgment}

We would like to thank Alessandro La Mela for his support in preparing the ground truth data for our experiments.

\bibliographystyle{IEEEtran} 
\bibliography{IEEEabrv,main}

\end{document}